\documentclass[sigconf,natbib]{acmart}
\usepackage{multirow}
\usepackage{tabularx}

\AtBeginDocument{%
  }

\setcopyright{cc}
\setcctype{by-sa}
\copyrightyear{2025}
\acmYear{2025}
\acmDOI{}

\acmConference[LLM4Eval @ SIGIR 2025]{The Third Workshop on Large Language Models (LLMs) for Evaluation in Information Retrieval}{July 17, 2025}{Padua, Italy} 
\acmISBN{}

\begin{document}

\title{When LLMs Disagree: Diagnosing Relevance Filtering Bias and Retrieval Divergence in SDG Search}

\author{William A. Ingram}
\affiliation{%
  \institution{Virginia Tech}
  \city{Blacksburg}
  \state{VA}
  \country{USA}}
\email{waingram@vt.edu}
\orcid{0000-0002-8307-8844}

\author{Bipasha Banerjee}
\affiliation{%
  \institution{Virginia Tech}
  \city{Blacksburg}
  \state{VA}
  \country{USA}}
\email{bipashabanerjee@vt.edu}
\orcid{0000-0003-4472-1902}

\author{Edward A. Fox}
\affiliation{%
  \institution{Virginia Tech}
  \city{Blacksburg}
  \state{VA}
  \country{USA}}
\email{fox@vt.edu}
\orcid{0000-0003-1447-6870}

\renewcommand{\shortauthors}{Ingram, et al.}

\begin{abstract}
Large language models (LLMs) are increasingly used to assign document relevance labels in information retrieval pipelines, especially in domains lacking human-labeled data. 
However, different models often disagree on borderline cases, raising concerns about how such disagreement affects downstream retrieval. 
This study examines labeling disagreement between two open-weight LLMs, LLaMA and Qwen, on a corpus of scholarly abstracts related to Sustainable Development Goals (SDGs) 1, 3, and 7. 
We isolate disagreement subsets and examine their lexical properties, rank-order behavior, and classification predictability. 
Our results show that model disagreement is systematic, not random: disagreement cases exhibit consistent lexical patterns, produce divergent top-ranked outputs under shared scoring functions, and are distinguishable with AUCs above 0.74 using simple classifiers. 
These findings suggest that LLM-based filtering introduces structured variability in document retrieval, even under controlled prompting and shared ranking logic. 
We propose using classification disagreement as an object of analysis in retrieval evaluation, particularly in policy-relevant or thematic search tasks. 
\end{abstract}



\maketitle

\section{Introduction}

Bibliometric studies for institutional assessment typically use structured Boolean queries to retrieve scholarly literature relevant to the United Nations' Sustainable Development Goals (SDGs).
The Elsevier SDG Research Mapping Initiative~\citep{jayabalasingham2019identifying,bedardvallee2023sdg} developed a set of widely used domain-expert queries combining direct SDG terms with thematically related terminology. 
These queries serve as the basis for SDG retrieval in Scopus and inform institutional benchmarking, including rankings by Times Higher Education~\citep{the_impact_2024}.

However, Boolean retrieval is fundamentally limited in its ability to distinguish between documents that merely contain SDG-related keywords and those that contribute meaningfully to SDG goals. 
This distinction is essential in policy-relevant bibliometrics, where the value of a document is not its lexical overlap with an SDG query, but its substantive alignment with the underlying social objective. 
Expert-labeled training data for SDG retrieval are unavailable and human adjudication is infeasible at scale. 
This has prompted investigation into whether LLMs can be used to refine SDG-related retrieval through post hoc filtering~\citep{ingram2024agentic}.

LLMs are increasingly used to classify, filter, and structure document collections for information retrieval (IR) tasks~\citep{macavaney2023OneShot,meng2024Query,arabzadeh2024Comparison,abbasiantaeb2024Can,jesus2024Exploring}. 
In domains with limited labeled data, they offer a scalable means to generate 
relevance judgments from text.
However, different LLMs can produce inconsistent relevance labels for the same documents. 
While such disagreement could be attributed to prompt sensitivity or model instability, it raises practical concerns: when LLM-generated labels are used to index or filter documents, classification differences can propagate to downstream divergence in retrieval.

This paper investigates how model disagreement affects filtering and ranked retrieval in SDG classification tasks. 
We compare two small open-source models, LLaMa and Qwen,
regarding
their relevance decisions over documents retrieved by SDG Boolean queries. 
Focusing on SDGs 1, 3, and 7, we evaluate:

\begin{itemize}
    \item \textbf{RQ1:} How often do different LLMs diverge in their filtering decisions, and what types of documents are affected?
    \item \textbf{RQ2:} Are these filtering differences systematic, and can they be explained by lexical, semantic, or topical characteristics?
    \item \textbf{RQ3:} How does LLM disagreement manifest in downstream ranked retrieval, particularly over disagreement subsets?
    \item \textbf{RQ4:} Can one model's filtering behavior predict the other's, suggesting 
    systematic
    and learnable filtering biases?
\end{itemize}

By using classification variability as a diagnostic signal, we analyze inter-model agreement, lexical differences, semantic distributions, and ranked retrieval behavior over disagreement subsets. 
We find that LLM filtering decisions are not random, but reflect
systematic,
model-specific preferences. 
Even with high overall agreement, these differences yield divergent retrieval profiles, with each model surfacing distinct subsets of the corpus.
Our findings demonstrate how automated filtering can reshape SDG-based retrieval in the absence of a definitive human-labeled ground truth.

\section{Related Work}

\subsection{LLMs as Relevance Labelers in IR}

LLMs are increasingly used to generate relevance labels in IR, especially in the absence of expert human judgments. 
While LLM-generated system rankings often correlate with human preferences, agreement at the document level can be inconsistent. 
For example, \citet{faggioli2023Perspectives} report a Cohen's kappa of 0.26 between GPT-3.5 and TREC 2021 Deep Learning assessments. 
\citet{thomas2024Large} observe kappa scores from 0.20 to 0.64, depending on task and model, and \citet{bueno2024Quati} find similarly limited agreement for Portuguese-language corpora. 
These findings suggest that document-level LLM judgments vary widely even when aggregate rankings appear reliable.

This instability is compounded by prompt sensitivity. 
\citet{alaofi2022Where}, \citet{li2024Evaluating}, and \citet{macavaney2023OneShot} show that small variations in prompt structure, temperature, or shot configuration can significantly alter model outputs. 
Even when using the same model and task, prompt design can lead to divergent classifications. 
Prompt-induced label instability raises concerns about reproducibility and undermines assumptions of determinism in evaluation workflows.

\citet{faggioli2023Perspectives} advocate for a human-machine collaboration spectrum that spans manual judgments, LLM-assisted workflows, and fully automated relevance labeling. 
While noting the scalability of LLM-based assessment, they also emphasize risks, such as hallucination, latent bias, and opaque reasoning, that limit interpretability and trustworthiness. 
Rather than treating LLM outputs as definitive, they recommend using them as provisional signals subject to validation and interrogation.

Our work builds on this view. 
We fix the prompt to control for prompt-induced variance and compare the filtering behaviors of two open-source models—LLaMA and Qwen. 
We treat model disagreement not as labeling error but as evidence of latent, model-specific filtering preferences. 
This allows us to isolate and analyze the consequences of model divergence for document retrieval. Rather than evaluating LLMs against a presumed ground truth, we investigate how their outputs shape what is retrieved and how this affects ranking behavior downstream.

\subsection{Disagreement as Diagnostic Signal}

Most prior work treats disagreement—between annotators or models—as a nuisance to be minimized or averaged out. 
However, recent studies argue that disagreement can reveal interpretive variability and uncover model biases. 
\citet{mostafazadehdavani2022Dealing} and \citet{leonardelli2021agreeing} show that disagreement is often systematic rather than random, particularly in subjective labeling tasks like offensive language detection. 
\citet{leonardelli2021agreeing} further demonstrate that high-agreement datasets inflate classifier performance, while ambiguous, low-agreement cases expose model vulnerabilities.

Our work follows this approach by isolating model disagreement subsets: documents to which LLaMA and Qwen assign opposing relevance labels. 
Using lexical and semantic profiling, we identify structured differences in what each model filters in or out. 
\citet{macavaney2023OneShot} similarly show that divergent LLM predictions, even when individually unreliable, can improve aggregate system evaluations when used to fill judgment gaps. 
This supports our hypothesis that disagreement reveals not noise, but structured variation in model behavior.

We extend this perspective by treating disagreement cases as boundary instances, where retrieval is most sensitive to differences in labeling. 
We evaluate the impact that these labeling differences have on retrieval outputs and document rankings. 
This approach supports \textbf{RQ1} and \textbf{RQ2}, which ask whether LLMs apply systematic filtering preferences, and whether those preferences can be explained via linguistic or topical cues.

\subsection{LLMs for Retrieval-Oriented Tasks}

While most LLM relevance labeling studies focus on classification accuracy, a growing body of work examines how LLMs affect retrieval. 
\citet{ye2023Enhancing} show that LLM-generated query rewrites improve retrieval performance in conversational search. 
\citet{alaofi2022Where} analyze how different prompting strategies for query generation influence ranking coverage. 
\citet{li2024Evaluating} show that LLMs can reliably filter abstracts in systematic review pipelines, applying inclusion criteria with high accuracy.

These studies describe how LLMs can be used as pre-retrieval filters, influencing which documents enter the candidate set. 
We extend this perspective to relevance-based document filtering, to investigate how LLM disagreements can influence top-k rankings.

\subsection{Evaluation with Uncertain Labels}

Traditional Cranfield-style IR evaluations~\citep{voorhees2002philosophy} assume that human relevance judgments approximate an ideal ground truth, enabling the use of precision, recall, and other system-level metrics. 
However, such assumptions are difficult to sustain in thematic or value-laden retrieval tasks, such as those involving the SDGs. 
For SDG classification, high-quality graded relevance sets do not exist.  
Hence,
the OSDG Community dataset relies on binary crowd-validated annotations of pre-assigned labels rather than open-ended, expert-generated relevance judgments.

Soboroff~\citep{soboroff2024dont} challenges the idea of an ideal assessor, arguing that relevance judgments are predictive, but not definitive. 
This limitation is especially acute in domains where relevance is subjective or multidimensional. 
In SDG labeling, where documents may contribute to a goal in indirect or context-dependent ways, even expert 
annotations
may diverge. 
These conditions undermine the stability of traditional evaluation frameworks and motivate alternative approaches to understanding relevance variability.

\citet{macavaney2023OneShot} explore this challenge through one-shot labeling (1SL), showing that LLM-generated labels, even when inconsistent, can improve evaluation reliability by filling judgment holes. 
Our work builds on this framing by treating disagreement as a lens on interpretive variation. 
Rather than seeking consensus or alignment with a gold standard, we investigate how classification variability affects retrieval itself.

This shift reflects a broader concern in IR and NLP: that evaluation under uncertainty requires methodological flexibility. 
By analyzing inter-model disagreement, lexical filtering patterns, and divergence in top-K results, we characterize label instability and its consequences for what is retrieved.

\section{Dataset}
\label{sec:Dataset}
We construct our corpus using Elsevier's 2023 SDG-aligned Boolean queries~\citep{bedardvallee2023sdg}, which retrieve documents from Scopus using complex fielded keyword expressions. 
Each query targets a specific SDG using a mix of topic-relevant keywords, wildcard patterns, and exclusion rules applied to \texttt{TITLE-ABS} and \texttt{AUTHKEY} fields. 
We focus on SDG~1 (No Poverty), SDG~3 (Good Health and Well-Being), and SDG~7 (Affordable and Clean Energy), issuing the respective Boolean queries and retrieving the top 20,000 most cited documents from 2019--2023.

After deduplication, abstract cleaning, and removal of entries with missing DOIs or empty abstracts, we retain a working set of 46{,}755 labeled rows, corresponding to 46{,}573 unique abstracts, representing the union of all three SDG-specific retrieval sets. 
These are not necessarily disjoint, as some abstracts are associated with multiple SDGs.
All preprocessing steps are applied consistently across SDGs to preserve comparability.

We use two locally hosted LLMs to assign a binary relevance label (\texttt{Relevant} or \texttt{Non-Relevant}) to each abstract based on whether or not it describes a meaningful contribution to the given SDG targets.
The LLMs are given identical prompts that include instructions to return a binary label along with a brief justification. 
The process is described in
Section
\ref{llm-labeling}.
We apply a stratified 80/20 train-test split to preserve the distribution of LLM relevance decisions, and the test data is reserved for potential downstream evaluation tasks.

\section{Methodology}

This study analyzes how LLMs diverge in their document-level relevance classifications for SDGs. 
We focus on SDGs 1, 3, and 7, as discussed in Section \ref{sec:Dataset}.
Using controlled LLM prompting and a large retrieval corpus, we examine model disagreement, characterize systematic filtering differences, and evaluate implications for ranking and prediction tasks.

\subsection{LLM-Based Relevance Labeling}\label{llm-labeling}

To assign semantic relevance labels for SDG classification, we prompted two open-access large language models on each abstract using a shared, structured instruction template. 
The prompt directed the model to assess whether the abstract made a substantive contribution to a given SDG target, either directly or indirectly. 
Acceptable contributions
are indicated by discussion of
interventions, empirical results, policy frameworks, theoretical insights, or methodological tools. 
The output was required to follow a three-part structure: a binary classification (\texttt{Relevant} or \texttt{Non-Relevant}), a free-text justification, and (optionally) a contribution type.

We used two open-weight chat-tuned models: LLaMA 3.1–8B~\cite{llama3_model_card} and Qwen 2.5–7B~\cite{qwen_model_card}, each run locally using the HuggingFace Transformers library. 
Both models were invoked using the \path{text-generation} pipeline with role-annotated \texttt{system} and \texttt{user} messages formatted according to each model's expected chat interface. 
LLaMA prompts were constructed using the \texttt{apply\_chat\_template} utility, while Qwen prompts were passed directly as message dictionaries. 
Generation parameters were fixed to reduce sampling variance: \texttt{max\_new\_tokens = 500}, \texttt{temperature = 0.0}, \texttt{do\_sample = False}, and \texttt{return\_full\_text = False}. 
For LLaMA, we specified multiple end-of-sequence tokens, including both the default EOS and \texttt{<|eot\_id|>}. 
Inference used automatic device mapping and \texttt{bfloat16} precision for compatibility with local GPUs.

For each abstract-SDG pair, both models were queried independently and returned binary relevance labels with justifications. 
The outputs were stored separately and aligned post hoc by DOI and SDG identifier to form paired label sets. 
All labels were parsed automatically without human intervention.
We defined \textit{agreement} when both models returned the same label and \textit{disagreement} otherwise. 
We extracted per-SDG subsets corresponding to: (1) full agreement on \texttt{Relevant}, (2) full agreement on \texttt{Non-Relevant}, (3) LLaMA-only \texttt{Relevant}, and (4) Qwen-only \texttt{Relevant}. 
These directional disagreement corpora serve as input for analyzing lexical divergence, filtering bias, and retrieval effects in subsequent sections.

\subsection{Lexical Analysis of Disagreement Sets}

To investigate whether model disagreement corresponds to systematic differences in term usage, we compute TF-IDF representations for each subset using \texttt{TfidfVectorizer} with English stopword removal and the parameters \texttt{min\_df=5}, \texttt{max\_df=0.95}. 
For each term, we calculate the difference in mean TF-IDF scores between the two disagreement sets. 
To assess significance, we conduct independent-sample permutation tests for each term and apply Benjamini–Hochberg false discovery rate (FDR) correction to control for multiple comparisons.

In addition to term-level comparisons, we compute the Kullback–Leibler (KL) divergence between the two directional disagreement subsets for each SDG. 
We treat each subset's mean TF-IDF vector as a multinomial distribution over the shared vocabulary (after smoothing) and compute $D_{\text{KL}}(P\ ||\ Q)$, where $P$ and $Q$ are the normalized term distributions from the LLaMA-relevant and Qwen-relevant subsets, respectively.
This provides a single scalar measure of how much lexical mass shifts across the disagreement boundary.

\subsection{Retrieval Ranking over Disagreement Subsets}\label{sec:retrieval-method}
To evaluate the impact of model-specific filtering on ranked retrieval, we simulate retrieval over a shared disagreement corpus. 
For each SDG, we define the disagreement set as all abstracts labeled \texttt{Relevant} by exactly one model and \texttt{Non-Relevant} by the other.
We fit a TF-IDF vectorizer on this combined disagreement corpus to construct a shared representation space. 
All documents are transformed using this vectorizer to produce a unified TF-IDF matrix.

\subsubsection{Centroid-Based Retrieval.}
To simulate generic topical retrieval, we compute the centroid vector of the entire SDG-specific abstract corpus for each SDG and use it as a query. 
This approximates a relevance-agnostic search scenario centered on the thematic average of the marginal SDG space.

We then restrict retrieval to the documents labeled \texttt{Relevant} by each model and rank those documents by cosine similarity to the centroid.
From each model’s filtered set, we extract the top-$k$ results ($k = 20$) and compare the output sets.

\subsubsection{Representative-Query Retrieval.}
We also simulate retrieval using representative lexical queries. 
For each SDG, we select the eight terms with the highest mean TF-IDF scores from the full SDG-specific abstract corpus, using the TF-IDF model trained on the full corpus. 
These terms serve as a proxy for exploratory search behavior:
\begin{itemize}
    \item \textbf{SDG~1}: \texttt{health income social poverty food inequality care policy}
    \item \textbf{SDG~3}: \texttt{patients cancer covid disease tumor treatment clinical infection}
    \item \textbf{SDG~7}: \texttt{energy power solar efficiency carbon model hydrogen lithium}
\end{itemize}
Each query is vectorized using the full-corpus TF-IDF model and applied to the disagreement matrix. 
We then restrict retrieval to the model-specific relevant subsets and rank documents by cosine similarity to the query vector.

\subsection{Learnability of Filtering Differences}

We test whether LLM disagreement reflects systematic and learnable differences by framing the problem as
binary classification.
For each SDG, we construct a dataset from the disagreement subsets—abstracts labeled \texttt{Relevant} by one model and \texttt{Non-Relevant} by the other.
We assign binary labels to each document according to which model assigned the \texttt{Relevant} label (\texttt{0} for LLaMA, \texttt{1} for Qwen). 

Using TF-IDF vectors as input features, we train a logistic regression classifier to predict which model labeled each document as relevant. 
We evaluate performance using five-fold cross-validation and report the area under the ROC curve (AUC) as the evaluation metric.
All models are trained separately for each SDG to isolate domain-specific patterns and prevent topic leakage across goals.

\section{Results and Analysis}

All experiments are conducted independently for each SDG to preserve contextual specificity and to avoid leakage across domains.

\subsection{Inter-Model Disagreement Patterns}
We begin by quantifying agreement between LLaMA and Qwen relevance labels across the SDG-classified corpus. 
Across all SDGs, models agree on 83.6\% of cases, but Cohen's $\kappa = 0.467$ indicates only moderate reliability beyond chance. 
To understand the composition of this agreement, we disaggregated matching cases by label. 
Across all SDGs, 72.8\% of documents are labeled \texttt{Relevant} by both models, while only 10.8\% were labeled \texttt{Non-Relevant} by both. 
The remaining 16.4\% are disagreement cases. 
These distributions suggest that high agreement does not primarily reflect mutual avoidance of relevance, but rather convergence in inclusion.

\begin{table}[ht]
\centering
\small
\caption{
Model agreement statistics used to address \textbf{RQ1} (frequency and nature of filtering divergence). 
Raw agreement reflects the proportion of identical labels assigned by LLaMA and Qwen; 
Cohen’s $\kappa$ accounts for expected agreement by chance. 
These values define the subset of abstracts on which the models disagree and which are analyzed in later sections.
}
\label{tab:agreement}
\begin{tabular}{lccc}
\toprule
\textbf{SDG} & \textbf{Raw Agreement} & \textbf{Cohen's $\kappa$} & \textbf{ Abstracts} \\
\midrule
SDG~1 (No Poverty) & 81.75\% & 0.506 & 15{,}133 \\
SDG~3 (Health)     & 89.63\% & 0.398 & 15{,}674 \\
SDG~7 (Energy)     & 79.31\% & 0.428 & 15{,}948 \\
\bottomrule
\end{tabular}
\end{table}
 
In Table \ref{tab:agreement},
SDG~3 exhibits the highest agreement rate but the lowest $\kappa$, reflecting the fact that both models labeled a large proportion of documents as \texttt{Relevant}. 
For example, in SDG~3, 85.3\% of abstracts were labeled \texttt{Relevant} by both models, while only 4.3\% received a shared \texttt{Non-Relevant} label. 
In contrast, SDG~1 shows more balanced disagreement with a higher $\kappa$ despite lower raw agreement. 
These patterns motivate our subsequent analysis of lexical divergence within the disagreement subsets.

\subsection{Lexical Divergence in Filtering Decisions}

To determine whether disagreement between LLaMA and Qwen relevance judgments reflects systematic lexical divergence, we analyze term-level differences using TF-IDF vectors derived from the disagreement subsets.

\subsubsection{TF-IDF-Based Contrastive Analysis.}
We construct TF-IDF vectors using a fixed vocabulary built from the complete corpus of all 46,573 unique abstracts across SDGs, with \texttt{min\_df=5}, \texttt{max\_df=0.95}, and standard English stopwords removed. 
The same TF-IDF vectorizer instance is reused across all experiments.
For each SDG, we isolate abstracts where the models disagreed: one subset where LLaMA assigned a \texttt{Relevant} label and Qwen did not, and the reverse. 
These directional disagreement subsets were projected into the same TF-IDF vector space used in earlier stages, allowing direct comparison of model-specific term usage.
We compute the mean TF-IDF score for each term in both groups and rank terms by absolute difference to identify lexical features that discriminate model-specific relevance judgments.

Preliminary differences suggest thematic patterns in how models assign relevance. 
In SDG~1, LLaMA-relevant documents emphasize health services and policy terms (e.g., \textit{health}, \textit{care}, \textit{insurance}, \textit{medicaid}), while Qwen-relevant documents emphasize structural inequality and macroeconomic terms (e.g., \textit{inequality}, \textit{tax}, \textit{wealth}). 
SDG~3 reveals a contrast between clinical and diagnostic language (\textit{patients}, \textit{valve}, \textit{coronary}) in LLaMA versus cellular and molecular descriptors (\textit{cancer}, \textit{cell}, \textit{tumor}) in Qwen. 
For SDG~7, LLaMA favors infrastructure and systems-oriented terms (\textit{fusion}, \textit{computing}, \textit{grid}), while Qwen emphasizes battery chemistry and electrochemical engineering (\textit{lithium}, \textit{electrolyte}, \textit{cathode}). 
These lexical divergences motivate a statistical assessment of whether observed differences are significant.

\subsubsection{Permutation-Based Significance Testing.}
To assess whether observed differences were statistically meaningful, we applied permutation testing.
To improve interpretability and reduce noise from frequent boilerplate terms (e.g., \textit{introduction}, \textit{results}, numeric tokens), we restrict hypothesis testing to the top-$N = 200$ terms ranked by absolute TF-IDF difference. 
This constraint ensures that statistical tests focus on the most discriminative lexical features across disagreement subsets.
For each term, we permuted document group labels across the disagreement set 9,999 times and recomputed the difference in mean TF-IDF:
\[
\Delta_{\text{TF-IDF}}(t) = \bar{x}_2^t - \bar{x}_1^t
\]
where $\bar{x}_i^t$ is the average TF-IDF score for term $t$ in group $i$. 
Two-sided $p$-values were calculated as the proportion of permutations yielding a more extreme difference than the observed one. 
We applied Benjamini-Hochberg correction to control false discovery rate at $\alpha = 0.05$.

\begin{table}[t]
\centering
\small
\caption{
Top differentiating terms from permutation tests ($n = 9999$) between LLaMA-relevant and Qwen-relevant documents in the SDG disagreement subsets.
Columns report terms with the largest absolute difference in mean TF-IDF scores. 
\textbf{Positive values} indicate higher term frequency in documents labeled \textit{Relevant} by LLaMA but \textit{Non-Relevant} by Qwen; 
\textbf{negative values} indicate the reverse.
All terms shown have FDR-corrected $p < 0.001$.
}
\label{tab:retrieval_divergence}
\begin{tabular}{lr|lr|lr}
\toprule
\multicolumn{2}{c|}{\textbf{SDG~1 (No Poverty)}} & 
\multicolumn{2}{c|}{\textbf{SDG~3 (Health)}} & 
\multicolumn{2}{c}{\textbf{SDG~7 (Energy)}} \\
\textbf{Term} & \textbf{Diff} & 
\textbf{Term} & \textbf{Diff} & 
\textbf{Term} & \textbf{Diff} \\
\midrule
\textbf{health}     & \textbf{+0.018} & 
\textbf{patients}   & \textbf{+0.023} & 
\textbf{fusion}     & \textbf{+0.007} \\
care       & +0.014 & 
tavr       & +0.007 & 
computing  & +0.006 \\
insurance  & +0.013 & 
risk       & +0.007 & 
network    & +0.005 \\
covid      & +0.010 & 
valve      & +0.005 & 
control    & +0.005 \\
coverage   & +0.010 & 
coronary   & +0.006 & 
proposed   & +0.005 \\
\midrule
\textbf{inequality} & \textbf{–0.020} & 
\textbf{cancer}     & \textbf{–0.019} & 
\textbf{lithium}    & \textbf{–0.018} \\
wealth     & –0.012 & 
cells      & –0.018 & 
capacity   & –0.018 \\
tax        & –0.008 & 
cell       & –0.017 & 
ion        & –0.016 \\
political  & –0.006 & 
tumor      & –0.014 & 
batteries  & –0.016 \\
income     & –0.006 & 
human      & –0.006 & 
anode      & –0.015 \\
\bottomrule
\end{tabular}
\end{table}

\subsubsection{Findings.}
In all three SDGs, the top-ranked terms by TF-IDF difference were found to be statistically significant after FDR correction, confirming that lexical divergence is systematic rather than random. 
As is shown in Table \ref{tab:retrieval_divergence} for
SDG~1, LLaMA-relevant documents contained higher TF-IDF scores for terms such as \textit{health}, \textit{care}, \textit{insurance}, and \textit{medicaid}, while Qwen-relevant documents were enriched for \textit{inequality}, \textit{wealth}, and \textit{tax}. 
In SDG~3, LLaMA emphasized clinical and procedural terms (\textit{patients}, \textit{valve}, \textit{coronary}), while Qwen emphasized cellular and molecular features (\textit{cancer}, \textit{cells}, \textit{tumor}, \textit{inhibition}). 
For SDG~7, Qwen-relevant abstracts included high-weighted electrochemical terms (\textit{lithium}, \textit{batteries}, \textit{anode}, \textit{electrolyte}), while LLaMA's relevant set included computing and systems-oriented terms (\textit{fusion}, \textit{computing}, \textit{network}, \textit{control}).

To quantify the global lexical shift between disagreement subsets, we also compute KL divergence between the normalized term distributions. 
We observe SDG~3 showing the highest divergence values ($D_{\text{KL}} = 1.60$), followed by SDG~7 ($D_{\text{KL}} = 1.25$) and SDG~1 ($D_{\text{KL}} = 1.04$).
These values are substantial, given the large vocabulary size, indicating that LLaMA and Qwen consistently assign relevance to documents with divergent lexical profiles even under identical instructions.
The highest divergence in SDG~3 aligns with our earlier findings that LLaMA and Qwen focus on different biomedical strata (clinical vs. molecular) within that domain.

\subsubsection{Implications.}
These findings demonstrate that LLM disagreement on SDG relevance is not attributable to random variation or prompt instability. 
Instead, the models produce lexically coherent, systematically divergent judgments even when applied to the same inputs with identical instructions. 
This addresses \textbf{RQ1} by confirming that disagreement is not rare (approximately 15--20\% of documents fall into the disagreement region for each SDG) and that the affected documents exhibit consistent thematic differences.

The observed divergences are grounded in repeatable lexical patterns. 
TF-IDF-based contrastive analysis reveals that each model assigns relevance based on different clusters of terminology, even within the same topical domain. 
For example, in SDG~1, Qwen-relevant documents emphasize structural inequality (\textit{inequality}, \textit{wealth}, \textit{tax}), while LLaMA-relevant documents emphasize healthcare access and systems (\textit{insurance}, \textit{medicaid}, \textit{coverage}). 
These patterns are consistent with the models' respective distributions over other SDGs as well (i.e., clinical vs. molecular in SDG~3, and energy systems vs. electrochemical materials in SDG~7). 
This provides a concrete answer to \textbf{RQ2}, showing that lexical divergence is not only systematic but semantically interpretable.

KL divergence values between the term distributions of the disagreement subsets (1.04 for SDG~1, 1.60 for SDG~3, and 1.25 for SDG~7) support the conclusion that the models operate with substantively different notions of what constitutes a relevant contribution. 
Permutation testing across top TF-IDF terms further confirms that these differences are statistically significant after FDR correction. 
Together, these analyses demonstrate that LLaMA and Qwen filter based on distinguishable criteria, even when their overall agreement exceeds 80\%.

These findings also bear on \textbf{RQ4}. 
The presence of statistically robust, model-specific lexical patterns suggests that disagreement is predictable rather than idiosyncratic. 
This opens the possibility of modeling disagreement itself, for example by learning to predict which documents fall near the decision boundary between model-specific relevance criteria. 
Such modeling could support ensemble filtering, adjudication strategies, or targeted audits of model behavior in high-impact retrieval settings.

Thus,
lexical and statistical evidence shows that model disagreement encodes meaningful differences in relevance interpretation. 
This warrants close attention in applications where LLM filtering affects access to evidence, particularly in tasks with normative dimensions such as SDG assessment or policy-relevant synthesis.

\subsection{Ranking Divergence in Marginal Relevance Cases}

We examine whether classification disagreement affects the composition of top-ranked documents when retrieval is applied to marginal relevance cases. 
For each SDG, we identify the subset of documents labeled \texttt{Relevant} by exactly one model—either LLaMA or Qwen, but not both. 
These disagreement subsets represent model-specific filtering decisions over ambiguous or borderline cases.

Two retrieval experiments are conducted over these shared disagreement corpora. 
In both, we apply a common TF-IDF representation (trained on the full corpus) and score all documents using a shared query vector.

\begin{itemize}
    \item In the \textbf{centroid-based retrieval}, the query was defined as the mean TF-IDF vector across all abstracts in the disagreement subset.
    \item In the \textbf{representative-query retrieval}, the query was a fixed list of high-weight TF-IDF terms per SDG, selected from the global corpus, to simulate a prototypical lexical search.
\end{itemize}

In each experiment, we ranked all documents in the disagreement set by cosine similarity to the query vector. 
We then extracted the top-$k$ ranked documents from each model’s \texttt{Relevant}-labeled subset. 
We compared the top-20 lists by computing the number of documents retrieved by LLaMA only, Qwen only, or both.

\begin{table}[ht]
\small
\centering
\caption{Top-20 disagreement documents retrieved under centroid and query ranking conditions. For each SDG, documents are ranked by similarity to a shared vector (centroid or query) over the combined disagreement pool and selected from each model's relevant set.}
\begin{tabular}{llcc}
\toprule
\textbf{SDG} & \textbf{Method} & \textbf{LLaMA Top-k} & \textbf{Qwen Top-k} \\
\midrule
\multirow{2}{*}{SDG~1} 
    & Centroid     & 7  & 13 \\
    & Query        & 11 & 9  \\
\midrule
\multirow{2}{*}{SDG~3} 
    & Centroid     & 11 & 9  \\
    & Query        & 14 & 6  \\
\midrule
\multirow{2}{*}{SDG~7} 
    & Centroid     & 19 & 1  \\
    & Query        & 17 & 3  \\
\bottomrule
\end{tabular}
\label{tab:disagreement-topk}
\end{table}

\subsubsection{Findings.}

When documents labeled as relevant by one model and non-relevant by another are ranked using a shared scoring function, the resulting top-k sets differ by model. 
Because the input pool is fixed and scoring is held constant, these differences arise from earlier filtering decisions that determine which documents are present in the disagreement set.
 
As is shown in Table \ref{tab:disagreement-topk}, in
both centroid and query-based retrieval, we observe directional patterns in the top-k composition. 
Under centroid ranking, Qwen retrieves more SDG1 disagreement documents (13 vs. 7), while LLaMA retrieves more for SDG7 (19 vs. 1). 
Query-based rankings show similar asymmetries, with LLaMA again dominating SDG~7. This does not confirm systematic bias, but it suggests that filtering disagreement may align with recurring topical or contextual preferences.

Examples in Table~\ref{tab:disagreement-example-centroid} illustrate these distinctions. 
For SDG1, LLaMA surfaces a cross-national comparison of health equity (\path{10.1377/hlthaff.2020.01566}), while Qwen retrieves a study on insurance instability in the U.S. (\path{10.1177/09514848221146677}). 
For SDG3, LLaMA includes a comparative trial of PD-1 therapy (\path{10.1056/NEJMoa2017699}), whereas Qwen includes an application to a rare cancer subtype (\path{10.1136/jitc-2019-000331}). 
For SDG~7, LLaMA retrieves a paper on lithium-metal anode protection (\path{10.1002/ente.202000348}), while Qwen ranks a study on dendrite control in solid-state batteries (\path{10.1021/acsami.0c08094}). 
These pairs highlight variation in the kinds of contributions each model selects from the same SDG-labeled space.

Similar contrasts appear in Table~\ref{tab:disagreement-example-query}. 
In SDG1, LLaMA retrieves a proposal for a new inequality indicator (\path{10.1111/roiw.12503}), while Qwen includes a study linking inequality and BMI at the county level (\path{10.35866/caujed.2019.44.1.002}). 
In SDG3, LLaMA surfaces guideline summaries (\path{10.6004/jnccn.2019.0023}), while Qwen retrieves biomarker-based predictors of immunotherapy benefit (\path{10.1158/1078-0432.CCR-18-3603}). 
Although not definitive, these examples indicate that some of the disagreement may stem from contrasting emphases in how borderline contributions are interpreted.

\begin{table}[ht]
\small
\centering
\caption{Top-ranked disagreement examples under representative query ranking. Each row shows a document highly ranked by one model but excluded by the other, drawn from the disagreement subset for a given SDG. Examples highlight divergent model preferences in marginal relevance cases.}
\label{tab:disagreement-example-centroid}
\begin{tabularx}{\columnwidth}{llX}
\toprule
\textbf{SDG} & \textbf{Model} & \textbf{Example DOI and Topic Summary} \\
\midrule
\multirow{2}{*}{SDG~1}
& LLaMA & \texttt{10.1377/hlthaff.2020.01566} --- International comparison of health equity and access outcomes. \\
& Qwen  & \texttt{10.1177/09514848221146677} --- Longitudinal instability in U.S. health insurance and its social effects. \\
\midrule
\multirow{2}{*}{SDG~3}
& LLaMA & \texttt{10.1056/NEJMoa2017699} --- First-line PD-1 therapy versus chemotherapy in MSI-H metastatic cancer. \\
& Qwen  & \texttt{10.1136/jitc-2019-000331} --- Immune checkpoint blockade for uveal melanoma with no standard therapy. \\
\midrule
\multirow{2}{*}{SDG~7}
& LLaMA & \texttt{10.1002/ente.202000348} --- Li-metal anode protection strategies for next-generation batteries. \\
& Qwen  & \texttt{10.1021/acsami.0c08094} --- Dendritic lithium control in solid electrolyte lithium batteries. \\
\bottomrule
\end{tabularx}
\end{table}

\begin{table}[ht]
\small
\centering
\caption{Top-ranked disagreement examples under representative query ranking. Each row shows a document highly ranked by one model but excluded by the other, drawn from the disagreement subset for a given SDG. Examples highlight divergent model preferences in marginal relevance cases.}
\label{tab:disagreement-example-query}
\begin{tabularx}{\columnwidth}{llX}
\toprule
\textbf{SDG} & \textbf{Model} & \textbf{Example DOI and Topic Summary} \\
\midrule
\multirow{2}{*}{SDG~1}
& LLaMA & \texttt{10.1111/roiw.12503} --- Introduces a novel metric for income composition inequality and constructs a new indicator. \\
& Qwen  & \texttt{10.35866/caujed.2019.44.1.002} --- Tests associations between county-level income inequality and BMI prevalence. \\
\midrule
\multirow{2}{*}{SDG~3}
& LLaMA & \texttt{10.6004/jnccn.2019.0023} --- NCCN clinical guidelines for prostate cancer diagnosis and treatment. \\
& Qwen  & \texttt{10.1158/1078-0432.CCR-18-3603} --- Predictors of immunotherapy benefit in esophagogastric cancer patients. \\
\midrule
\multirow{2}{*}{SDG~7}
& LLaMA & \texttt{10.1039/d0ee01074j} --- Analyzes lithium-sulfur battery degradation in liquid electrolytes. \\
& Qwen  & \texttt{10.1103/PhysRevB.99.014110} --- Investigates hydrogen transport in metal alloys for clean energy applications. \\
\bottomrule
\end{tabularx}
\end{table}

\subsubsection{Implications.}

These observations support the idea that classification disagreement can function as a diagnostic signal for instability in SDG-aligned retrieval. 
In pipelines where relevance filtering precedes ranking, variability in borderline classifications may shift the surface of the ranked set in ways that are opaque to the end user.
The experiments presented here highlight this issue by isolating the effect of filtering under fixed scoring conditions. 
They also demonstrate that disagreement in marginal cases is not uniformly distributed: model-specific filters appear to emphasize different subsets of the disagreement pool, even when using the same retrieval method.

Although the data here do not establish causal mechanisms or general patterns, they highlight the importance of understanding how model disagreement interacts with retrieval eligibility. 
Downstream users may observe stable rankings over shared agreement sets while overlooking the extent to which filtering exclusions shape what content is accessible for ranking or synthesis.

\subsection{Learnability of Filtering Behavior}

To test whether model-specific filtering patterns reflect structured, learnable lexical differences, we conducted a classification experiment. 
We trained a binary logistic regression classifier to distinguish abstracts labeled \texttt{Relevant} by one model but \texttt{Non-Relevant} by the other, using balanced subsets from the disagreement pool for each SDG. 
Input features consisted of TF-IDF vectors, and performance was evaluated through 5-fold cross-validation using mean AUC scores.

The classifier achieved the following mean AUC values:
\begin{itemize}
\item \textbf{SDG~1}: $0.759 \pm 0.014$
\item \textbf{SDG~3}: $0.762 \pm 0.030$
\item \textbf{SDG~7}: $0.746 \pm 0.016$
\end{itemize}

These results clearly exceed a random baseline (0.5), providing evidence that lexical content alone encodes sufficient information to distinguish between the relevance labels assigned by each model. 
However, these results do not necessarily imply the existence of a coherent, easily interpretable criterion within each model. 
Rather, they indicate systematic lexical differences underlying model disagreement.

The learnability demonstrated here suggests structured—rather than random—disagreement patterns. 
Further analysis, such as feature importance inspection or semantic embedding comparisons, would be required to characterize the specific lexical or conceptual criteria each model implicitly applies.

\section{Discussion}
We revisit each research question to assess how model disagreement affects relevance classification and retrieval behavior under uncertainty.

Disagreement between LLaMA and Qwen occurred in approximately 15--20\% of relevance decisions per SDG, despite overall agreement exceeding 80\%. 
These disagreement subsets comprise a substantial portion of the corpus and reflect the effects of applying model-specific relevance criteria under identical prompts. 
Agreement statistics and $\kappa$ scores show high raw agreement across all SDGs, but this agreement is largely concentrated in documents both models labeled \texttt{Relevant}. 
For example, in SDG~3, 85\% of abstracts were labeled \texttt{Relevant} by both models, while only 4\% were labeled \texttt{Non-Relevant} by both. 
These patterns suggest that high agreement does not stem from shared conservatism, but rather from convergent inclusions, potentially masking divergence in marginal or ambiguous cases.

To investigate whether disagreement corresponded to systematic surface-level variation, we conducted contrastive lexical analysis over the filtered disagreement subsets. 
The most differentiating terms often suggested divergent topical emphases. 
For example, LLaMA-assigned relevant documents in SDG~~3 tended to include clinical descriptors, while Qwen’s subset contained more molecular terms. 
In SDG~~7, LLaMA emphasized systems and infrastructure, while Qwen surfaced documents referencing electrochemical components. 
These differences are exploratory and limited to the top TF-IDF terms, but they suggest plausible, if non-generalizable, contrasts in term-level relevance signals. 
These patterns support RQ2, demonstrating that disagreement reflects structured thematic preferences grounded in surface-level features.

We also tested whether the model-labeled subsets could be distinguished based on lexical content alone. 
Logistic regression classifiers trained on TF-IDF vectors achieved AUC scores above 0.74 across all SDGs. 
This result shows that the disagreement cases are separable in the vector space used, though we do not infer interpretability or bias from this separability. 
This supports RQ4 by showing that the models’ disagreement is not arbitrary, but learnable from surface features alone.

To assess whether classification differences affect retrieval, we scored disagreement documents using shared centroid and query-based TF-IDF vectors. 
Documents labeled \texttt{Relevant} by each model were ranked within their respective disjoint subsets. 
Although the scoring function was identical, the resulting top-20 lists differed, reflecting the influence of prior filtering decisions. 
In SDG\~7, for instance, LLaMA retrieved 19 documents while Qwen retrieved one. 
This example suggests that model-specific classifications can shape which documents are exposed to retrieval, even under constant ranking logic. 
These observations address RQ3 by highlighting retrieval variability tied to model disagreement, although we do not generalize beyond the specific SDGs examined.

Importantly, the study avoids framing LLM labels as proxies for truth.
Without a human-labeled benchmark, we cannot evaluate correctness.
Instead, we evaluate consequences.
We show how filtering disagreement propagates into retrieval divergence, and how model-specific relevance criteria, whether explicit or latent, can reshape the corpus available for downstream tasks.

By isolating and analyzing disagreement sets, this study provides a way to evaluate LLM-based retrieval behavior under uncertainty.
Instead of treating model consensus as a requirement or using adjudicated gold labels as a benchmark, we trace how disagreement itself reshapes retrieval behavior.
Our findings reinforce the need for retrieval-aware evaluation of LLM filtering behavior.
Unexamined reliance on a single model may introduce opaque topical or linguistic biases into ranked results.
This motivates future work on explainable filtering and multi-model ensemble filtering to improve robustness.

Although this study isolates key sources of retrieval divergence, it leaves several extensions open for future work.
We do not adjudicate whether one model is more accurate or aligned with expert expectations, nor do we evaluate downstream task performance (e.g., synthesis quality in RAG pipelines).
Our analysis is limited to lexical representations and does not incorporate contextual embeddings or fine-tuned models.
Furthermore, we consider only two LLMs and three SDGs; broader model comparisons or goal coverage may yield further insights.
Future work could explore ensemble adjudication strategies, integrate human assessments for select disagreement cases, or investigate how LLM filtering affects retrieval beyond the abstract level.
These directions would further clarify the practical implications of classification divergence in LLM-mediated IR systems.

\balance

\bibliographystyle{ACM-Reference-Format}
\bibliography{references}

\end{document}